\documentclass[10pt,twocolumn]{revtex4-1}
\usepackage{amsmath,amssymb}
\usepackage{graphicx}
\usepackage{dcolumn}
\usepackage{bm}

\usepackage[usenames,dvipsnames]{color}

\begin{document}

\title{Unified model of hyperthermia via hysteresis heating in systems of interacting magnetic nanoparticles}

\author{S. Ruta}
\email{sir503@york.ac.uk}
\affiliation{Department of Physics, University of York, Heslington, York YO10 5DD, United Kingdom}
\author{O. Hovorka}
\affiliation{Faculty of Engineering and the Environment, University of Southampton, Highfield, Southampton, United Kingdom}

\author{R. Chantrell}
\affiliation{Department of Physics, University of York, Heslington, York YO10 5DD, United Kingdom}

\begin{abstract}

We present a general study of frequency and magnetic field dependence of the specific heat power produced during field-driven hysteresis cycles in magnetic nanoparticles with relevance to hyperthermia applications in biomedicine. Employing a kinetic Monte-Carlo method with natural time scales allows us to go beyond the assumptions of small driving field amplitudes and negligible inter-particle interactions, which are fundamental to applicability of the standard approach based on linear response theory. The method captures the superparamagnetic and fully hysteretic regimes and  the transition between them.  Our results reveal unexpected dipolar interaction-induced enhancement or suppression of the specific heat power, dependent on the intrinsic statistical properties of particles, which cannot be accounted for by the standard theory. Although the actual heating power is difficult to predict because of the effects of interactions, optimum heating is in the transition region between the superparamagnetic and fully hysteretic regimes.

\end{abstract}

\maketitle

Understanding the field-driven thermally activated processes in magnetic nanoparticle (MNP) systems has been at the cornerstone of developments of applications in magnetic recording and biomedicine\cite{Search,Laurent2008,Colombo2012}. Thermal fluctuations not only determine a criterion for stability of the magnetic state (e.g. information bit) but also introduce new intrinsic time scales competing with those of external driving forces. A consistent out-of-equilibrium thermodynamic theory applicable to nanoparticle systems in this regime has not yet been developed, which radically complicates the design and optimization procedures in practice. For example, quantifying the specific heating rates in systems of MNPs with statistically distributed properties, such as those envisaged for `heat assisted' (hyperthermia) cancer treatment methodologies {\cite{Martinez-Boubeta2013,Hergt2008,Mamiya2011,Branquinho2013}, remains an unsolved issue.  Although the mechanisms giving rise to heating are not fully understood, it has been suggested  that hysteresis may be the dominant mechanism \cite{Branquinho2013, Suto2009a, Verde2012}, and it is this mechanism which is considered here.

Generally, if the characteristic time scales of thermal fluctuation modes ($\tau$) are long, exceeding those of fast intrinsic dynamical processes and the external field period (inverse of frequency, $f^{-1}$), a typical observation of field-dependent magnetization $M(H)$ is the rate-independent (static) hysteresis loop. It arises in the presence of a large number of metastable states emerging from volume and anisotropy dispersions in MNPs and the effects of inter-particle interactions. So far, self-consistent mathematically tractable calculations of the specific heat power (${\cal P}$) in this limit have been only based on simplified Preisach models \cite{Mayergoyz1991}. 
At the opposite extreme of time scales, $\tau<<f^{-1}$, a system of MNPs is in thermodynamic equilibrium during the experimental time window at a given field $H$, defining the  superparamagnetic regime \cite{Cullity2011}. This requires a fundamentally different theoretical description based on the equilibrium thermodynamics, the $M(H)$ takes the form of a Langevin function without hysteresis  \cite{Cullity2011}, and as a result ${\cal P}=0$.

At the crossover between the hysteretic and superparamagnetic regimes, i.e. when $\tau\sim f^{-1}$, direct calculations of ${\cal P}$ become much more difficult. Dynamical linear response theory\cite{zwanzig2001} has been applied to a non-interacting system of MNPs close to equilibrium \cite{Rosensweig2002}, hereafter referred to as Rosensweig theory (RT), which gives:
\begin{equation}\label{rosensweig}
 {\cal P} = \pi\mu_0\chi_0 H_0^2 f\frac{2\pi f\tau}{1+(2\pi f\tau)^2}
\end{equation}
where $\chi_0$ is the initial zero field equilibrium susceptibility, usually determined experimentally or calculated from the Langevin function, and $H_0$ is the amplitude of the sinusoidal external magnetic field of frequency $f$. Here the $\tau$ depends on the intrinsic fluctuation mechanisms, often taken to relate to Brownian and/or N\'eel relaxation \cite{Hergt2008}. In addition to RT leading to Eq.\eqref{rosensweig}, several useful phenomenological \cite{Vallejo-Fernandez2013a,Hergt2006} and empirical models \cite{Branquinho2013,Mamiya2011,Martinez-Boubeta2012}  have also been developed to evaluate ${\cal P}$. However, in these models and also in all of the aforementioned approaches, the interaction effects have either been ignored or are hard to access on a systematic basis. 

Many studies have focused on  determining the optimal conditions in terms of intrinsic properties  and their distribution (particle size, anisotropy value, easy axis orientation)\cite{Verde2012,Martinez-Boubeta2012,Hergt2006,Sohn2010,Vallejo-Fernandez2013a}, extrinsic properties (AC field amplitude, AC field frequency)\cite{Mehdaoui2011,Serantes2010,Reeves2014} and the role of dipole interactions \cite{Martinez-Boubeta2013,Serantes2014,Martinez-Boubeta2012,Landi2014}, and have stressed the large influence of the distribution of magnetic properties. 
Also, most studies have concentrated on small field amplitude and further extensions will require going beyond the RT.
Martinez-Boubeta \cite{Martinez-Boubeta2012} and co-workers proposed a theoretical model, valid in RT regime, including inter-particle interaction  for systems of chains. To study the heating rates as function of field amplitude other authors  used the standard Metropolis MC method for system of identical particles \cite{Martinez-Boubeta2013,Serantes2014}. However it is important to note that the standard MC approach suffers from the difficulty of precise time quantification of the MC step.  Landi used a mean field approach for the dipole interactions\cite{Landi2014}.

Full understanding of the hyperthermia heating process requires a model incorporating these key elements: 1) it must span the full range of applicability beyond the linear response regime, 2) include the effects of inter-granular interactions, and 3) be fully time-quantified to allow frequency dependent studies. Although individual investigations incorporate some of these factors, a generalised model unifying all the key factors for hyperthermia is needed if hyperthermia is to be developed and optimised as a therapy. 

The present work incorporates the effects of inter-particle interactions within a self-consistent kinetic Monte-Carlo (kMC) modelling framework\cite{Chantrell2000}. The kMC model naturally includes the time scales of intrinsic thermal fluctuations without any further need for calibration, as opposed to Time-Quantified Metropolis Monte-Carlo methods \cite{Nowak2000}, and as we will show, genuinely reproduces the superparamagnetic regime, the `metastable' and `dynamic' hysteresis mechanisms \cite{dynvsmeta}, and their frequency dependent crossover. This allows for systematic validation of the standard theories, such as RT, and establishing their range of validity in practical applications where interactions cannot be ignored. Furthermore, our approach allows us to quantify the importance of metastability in the heat production in MNP assemblies and thereby gauge its significance with respect to the superparamagnetic heating anticipated according to RT and Eq. \eqref{rosensweig}.

\section{Results}

\subsection{Kinetic Monte-Carlo model}
The kinetic Monte-Carlo (kMC) method (see the Methods section for details) systematically incorporates the complexity of realistic particle distributions, thermal fluctuations, and time varying external fields (Fig. \ref{fig12}). In the present work, the uni-axial anisotropy $\vec k_i = K_i\hat k_i$, particle diameter $D_i$, and particle positions are randomized as illustrated in Fig. \ref{fig12}(c-d). The importance of interactions is controlled by adjusting the particle packing fraction $\epsilon$, with $\epsilon=0$ implying the non-interacting case. The system energy is: 
\begin{equation}\label{energy}
E = \sum_i(K_iV_i(\vec k_i\times\hat m_i)^2-V_iM_s\hat m_i\cdot(\vec H+\vec H_i^\textrm{dip}))
\end{equation}
with $V_i$ being the volume of a particle $i$, $M_s$ the saturation magnetization, $\hat m_i$ the particle moment normalized to unity. The effective local field acting on particle $i$ is given by the sum of the external applied field, $H$, and the interaction field, $H_i^\textrm{dip}$.
\begin{figure}[!t]
\begin{center}
\hspace{0.04\textwidth}
\includegraphics[angle = -90,width = 8.5cm]{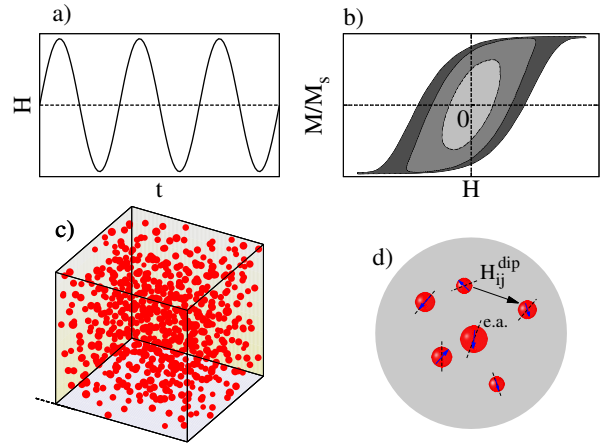}
\caption{Schematic illustration of key elements in the kinetic Monte-Carlo modelling framework to reproduce realistic experimental conditions. a) Sinusoidal external driving magnetic field magnetising the particle system and producing b) dynamical or metastable hysteresis loop depending on the applied field frequency. c-d) Typical distribution of particles of varying positions, size, anisotropy, and controllable packing fraction allowing to tune dipolar interactions between the particles.
} 
\label{fig12}
\end{center}
\end{figure}
It can be shown, that this formulation of the kMC approach \cite{Hovorka2014} provides a direct solution method for the master-equation:
\begin{equation}\label{master}
\frac{\partial M(t)}{\partial t} = \tau^{-1}(M_0(t)-M(t)).
\end{equation}
where $\tau$ is a  relaxation time constant dependent non-trivially on the intrinsic relaxation times of individual particles $\tau_i$ (see the Methods section). Eq. \eqref{master} is precisely the master-equation used to originally derive Eq. \eqref{rosensweig} under the assumptions of spherical random distribution of $\vec k_i$ in Eq. \eqref{energy} and the absence of inter-particle interactions. Hence, Eqs. \eqref{energy} and \eqref{master} establish an important link between the kMC and the RT approach. 

\begin{figure*}[!ht]
\begin{center}
\hspace{0.04\textwidth}
\includegraphics[angle = -90, width=\linewidth]{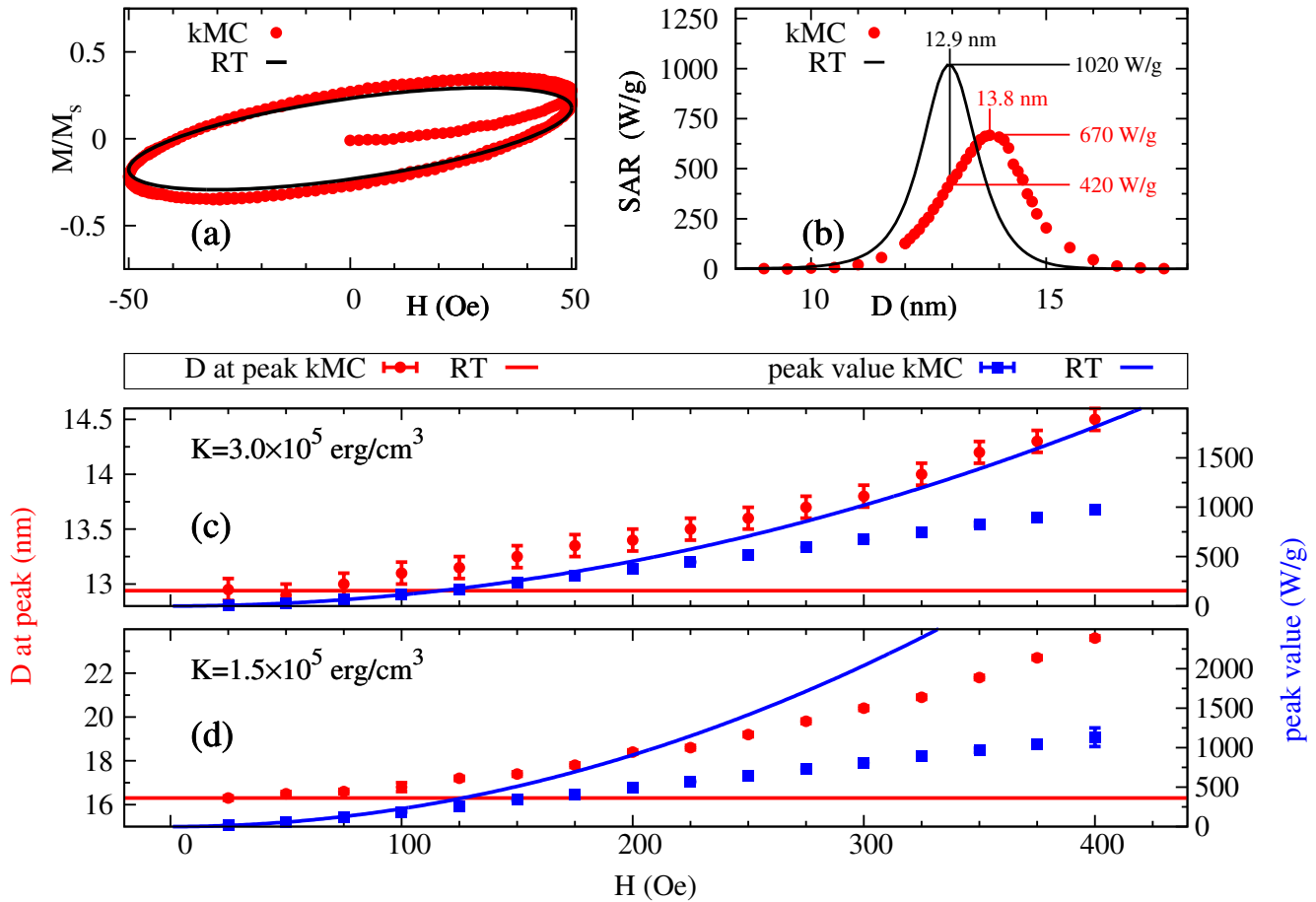}
\caption{Comparisons between RT (solid lines) and kMC calculation (symbols) for a system of identical particles with easy axes aligned in the field direction for: T=300K, $M_s$=400 $emu/cm^3$, K=3 $ \cdot 10^5 erg/cm^3$, f =100KHz. (a-b): Magnetisation loops for particle size of 13 nm at 50 Oe (a) and SAR as function of particle size at 300 Oe (b). (c-d): Maximum SAR value and the location of the peak in SAR  for K=3 $ \times 10^5 erg/cm^3$ (c) and K=1.5 $ \times 10^5 erg/cm^3$ (d).}
\label{fig5}
\end{center}
\end{figure*}
Throughout the study we consider a system of 1000 particles, and choose a parameter set typical for hyperthermia applications, specifically a sinusoidal applied field with $f$ = 100 kHz and amplitude $H_{0} < 600$ Oe, such that the product $H_{0}f$ is below the biological discomfort level of $\approx 6 \times 10^7$ Oe/s \cite{Hergt2007}. The particle temperature is set to $T=300$ K, thus ignoring the self-heating effect, which is equivalent to assuming infinite heat capacity of particles. We will furthermore assume variable distributions of anisotropy constant $\vec k_i$ and distributions of particle positions and volumes, which will allow control of the particle volumetric packing fraction $\epsilon$ and thus the importance of dipolar interactions. By definition, $\epsilon = 0.0$ corresponds to a non-interacting particle system. The saturation magnetisation $M_s$ will be initially set to 400 emu/cm$^3$ (i.e. magnetite like particles), and later allowed to vary to control the dipolar interaction strength further. The importance of dipolar interactions is expected to scale with the ratio of the local dipolar energy from the nearest neighbour particles to the anisotropy energy, i.e. $ \zeta = M_{s}^{2}\epsilon/K$ (see the Methods section).

\subsection{Ideal case: comparison of kMC with RT for system of identical particles}
Next we use the introduced kMC model to investigate the dynamics of combined superparamagnetic and hysteretic heating mechanisms. 
To establish and demonstrate the link between kMC and RT it is useful to first consider a simplified system of identical particles having anisotropy axes $\hat k_i$ oriented parallel and along the applied field direction. To quantify the heating produced by the nanoparticles we will use the quantity called specific absorption rate (SAR), defined as the heat dissipation $\cal P$ per unit mass (see the Methods section).

The calculated behaviour in the RT regime and beyond is shown in Fig. \ref{fig5} assuming $K_i=3 \times 10^5$ $erg/cm^3$ for all particles $i$.
The solid lines in Fig. \ref{fig5} show a comparison between standard calculations based on Eq.\eqref{rosensweig} and the kMC predictions. 
Consider first the behaviour for a  maximum applied field amplitude sufficiently small that the system is within the linear response regime and a direct comparison with RT is valid. 
For a non-interacting system both RT and kMC models predict  the same optimal value of SAR. 
In Fig. \ref{fig5}(a) we show  magnetization cycles starting from an initially demagnetized state.
It can be seen from Fig. \ref{fig5}(a) that the numerical results are in good agreement with RT. The rather elliptical loops are typical of dynamic hysteresis in systems in transition from truly thermal equilibrium to fully hysteretic behaviour. Essentially, although the behaviour of the particles can be characterised as superparamagnetic from quasi-static measurements, the heating arises from the onset of dynamic hysteresis at high frequencies, which is also to be associated with superparamagnetism, as opposed to the metastable hysteresis to be discussed below.

\begin{figure}[t!]
\begin{center}
\hspace{0.04\textwidth}
\includegraphics[angle = -90,width = 8.5cm]{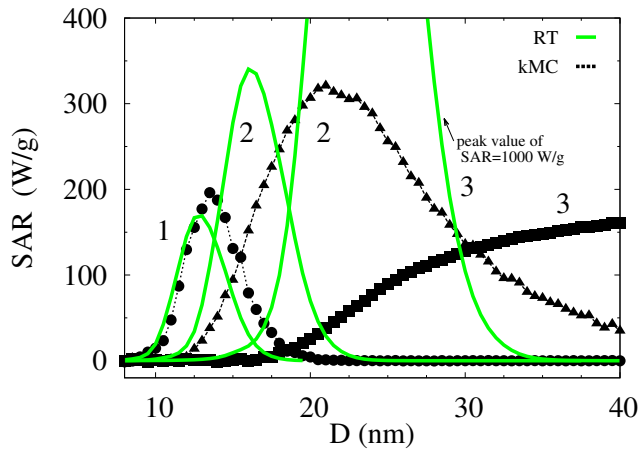}
\caption{Kinetic Monte-Carlo calculations of SAR as a function of the mean particle size $D$ assuming non-interacting system, in comparison with the predictions based on the RT (solid green lines). Considered are three values of mean anisotropy constant $K$: $3 \times 10^5$ erg/cm$^3$ (circles, curve set 1 peaking at low $D$),  $1.5 \times 10^5$ erg/cm$^3$ (triangles, curve set 2 peaking in the intermediate $D$ range), and $0.5 \times 10^5$ erg/cm$^3$ (squares, curve set 3 saturating for large $D$). Calculations correspond to log-normal distributions of size and anisotropy constants both with standard deviations 0.1, spherically distributed anisotropy axes, and field amplitude $H_0 = 300$ Oe.}
\label{fig10}
\end{center}
\end{figure}
Next we consider larger field amplitude beyond the linear response regime while the magnitude is constrained  to remain within the biological discomfort limit as introduced earlier. To determine the optimal particle size for a given field we have calculated SAR values as a function of mean particle diameter ($D$). The results are given in Fig. \ref{fig5}(b) for a field amplitude of $H_0=300$ Oe calculated using both RT and our kMC model. At this field the system is in a transition region between the linear approximation and fully hysteretic behaviour, which results in pronounced deviations between the two modelling approaches. In particular, the RT considerably overestimates the peak in SAR (by $\sim$ 50\%) and underestimates the corresponding optimum particle size. As a result, when using RT, the predictions for optimum particle size based on the peak in SAR are inaccurate and highly sensitive to uncertainties.

This trend is further examined in Fig. \ref{fig5}(c), where we systematically investigate the dependence of the optimal particle size and the maximum value of SAR on the field amplitude $H_0$. This is determined for every $H_0$ value by first finding the peak in the SAR as a function of $D$ dependence, similar to the case illustrated in Fig. \ref{fig5}(b). 
For small $H_0$ the system is in the linear response regime and both the peak value of SAR and the peak position as calculated from the kMC approach remain in agreement with the RT. As the $H_0$ increases, the RT clearly begins to underestimate the optimum particle diameter that would ideally correspond to the peak in SAR. Also, the dependence of the peak value on the field $H_0$ is close to quadratic (in agreement with RT) in small fields but becomes closer to linear with increasing $H_0$.
Interestingly, decreasing the value of the uniaxial anisotropy constant, $K$, magnifies the non-linearity of the magnetic behaviour which, as shown in Fig. \ref{fig5}(d), leads to more significant variation of the peak position. As the field amplitude is limited by the biological discomfort limit, we will keep $H_0$=300 Oe and vary K to investigate the non-linear regime.

\subsection{Non-linear regime at the crossover between superparamagnetism and metastable hysteresis in distributed particle assemblies}
The above considerations based on assuming an idealised system of identical non-interacting particles clearly demonstrate that as the field amplitude $H_0$ increases the RT begins to significantly deviate from the equivalent benchmark kMC calculations on entering the non-linear regime. To explore the transition regime further requires including various sources of randomness as they may appear in practical applications. This includes distributions of particle diameter, $D_i$, and anisotropy constants, $K_i$, which will be assumed to take the form of realistic log-normal distributions with expectation values $D = \langle D_i\rangle$ and $K = \langle K_i\rangle$, respectively, and standard deviations set to 0.1 in both cases, and also spherically random distribution of anisotropy vectors $\hat k_i$. 

Calculations of the dependence of SAR  on $D$ for three values of $K$ varying in the range 0.5-3$\times 10^5$ erg/cm$^3$ are shown in Fig. \ref{fig10}. For high values of $K$ (curve set 1 in Fig. \ref{fig10}), the predictions of RT reproduce the kMC calculations, since the intrinsic switching fields of particles $H_K = 2K/M_s$ are too high in comparison to $H_0$, and thus the primary mechanism contributing to heating is of superparamagnetic origin leading to dynamical hysteresis loops. Increasing the particle diameter $D$ leads to the suppression of thermal activation due to the particle moment blocking, and as a result SAR reduces to zero. On the other hand, for low values of $K$ (curve set 3), the $H_K$ is reduced and particle switching driven by external field through the appearance and annihilation of metastable states becomes a dominant mechanism contributing to SAR. In this regime, the RT becomes inapplicable due to the significant metastable hysteresis. Increasing $D$ again leads to particle blocking and suppression of thermally activated behaviour, and as a result the SAR relates directly to the area of a static minor hysteresis loop limited by field amplitude $H_0$ which no longer depends on $D$, thus justifying the observed saturating trend.

The intermediate range of $K$ (curve set 2) marks the crossover range where both mechanisms play a role, leading to a  relatively broad peak for the intermediate anisotropy value. This is an important prediction since it is a major advantage from the experimental point of view, in that it reduces the need for close control of the particle mean size, while retaining a high SAR value. The overall predictions for the dependence of SAR on $D$ begin to systematically deviate from RT as the value of $K$ decreases, first due the nonlinearities leading to failure of the linear response theory, and thereafter due to the onset of metastable hysteresis.
These observations suggest that the parameter controlling the type of behaviour is the ratio $H_0/H_K$. For $H_0/H_K > 1$ the behaviour is dominated by the metastable hysteresis while for $H_0/H_K << 1$ the SAR is of superparamagnetic origin and quantifiable by RT.

\begin{figure*}[ht!]
\begin{minipage}{0.49\linewidth}
\includegraphics[angle = -90,width=\linewidth]{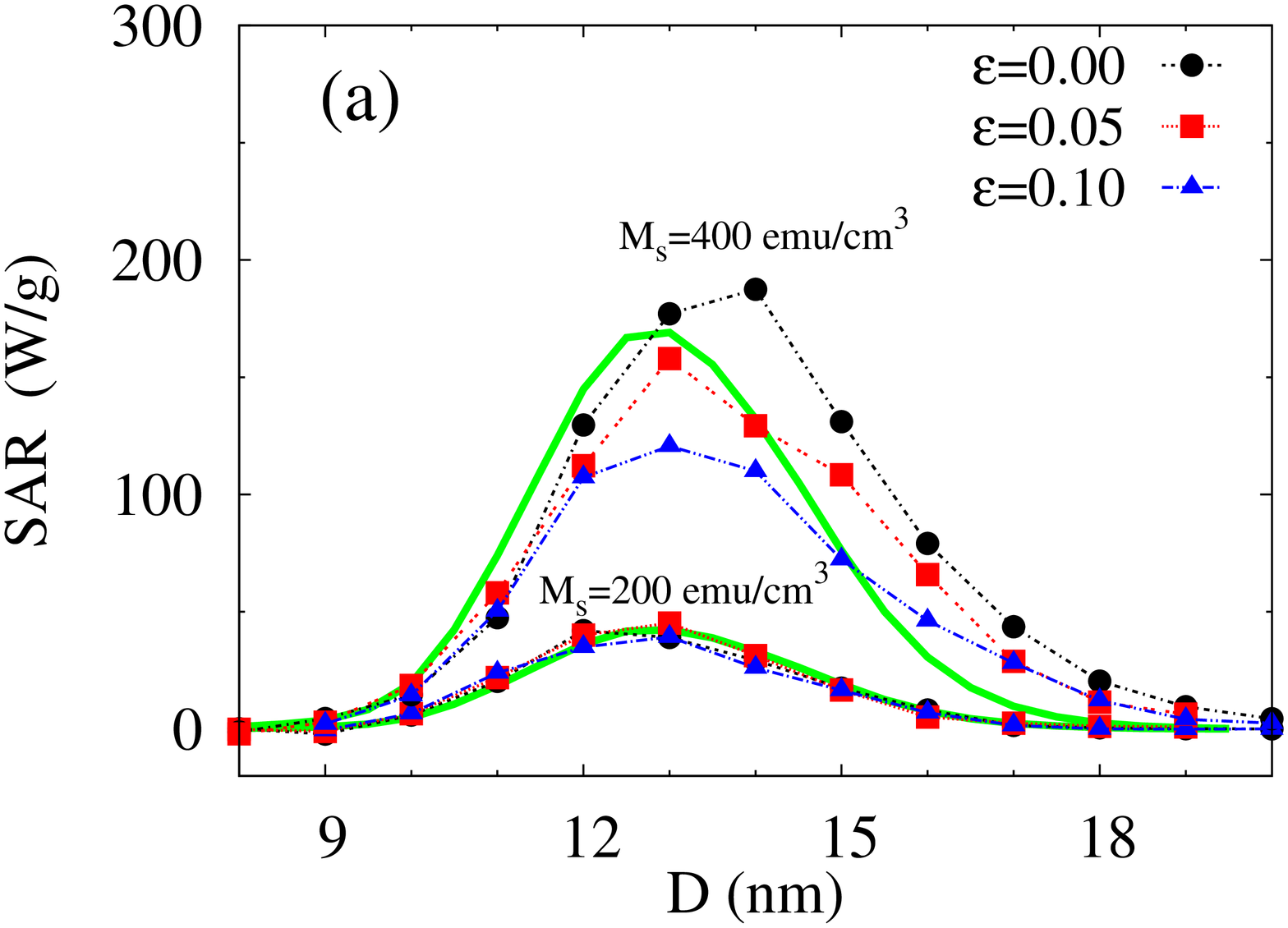}
\end{minipage}
\hfill
\begin{minipage}{0.49\linewidth}
\includegraphics[angle = -90,width=\linewidth]{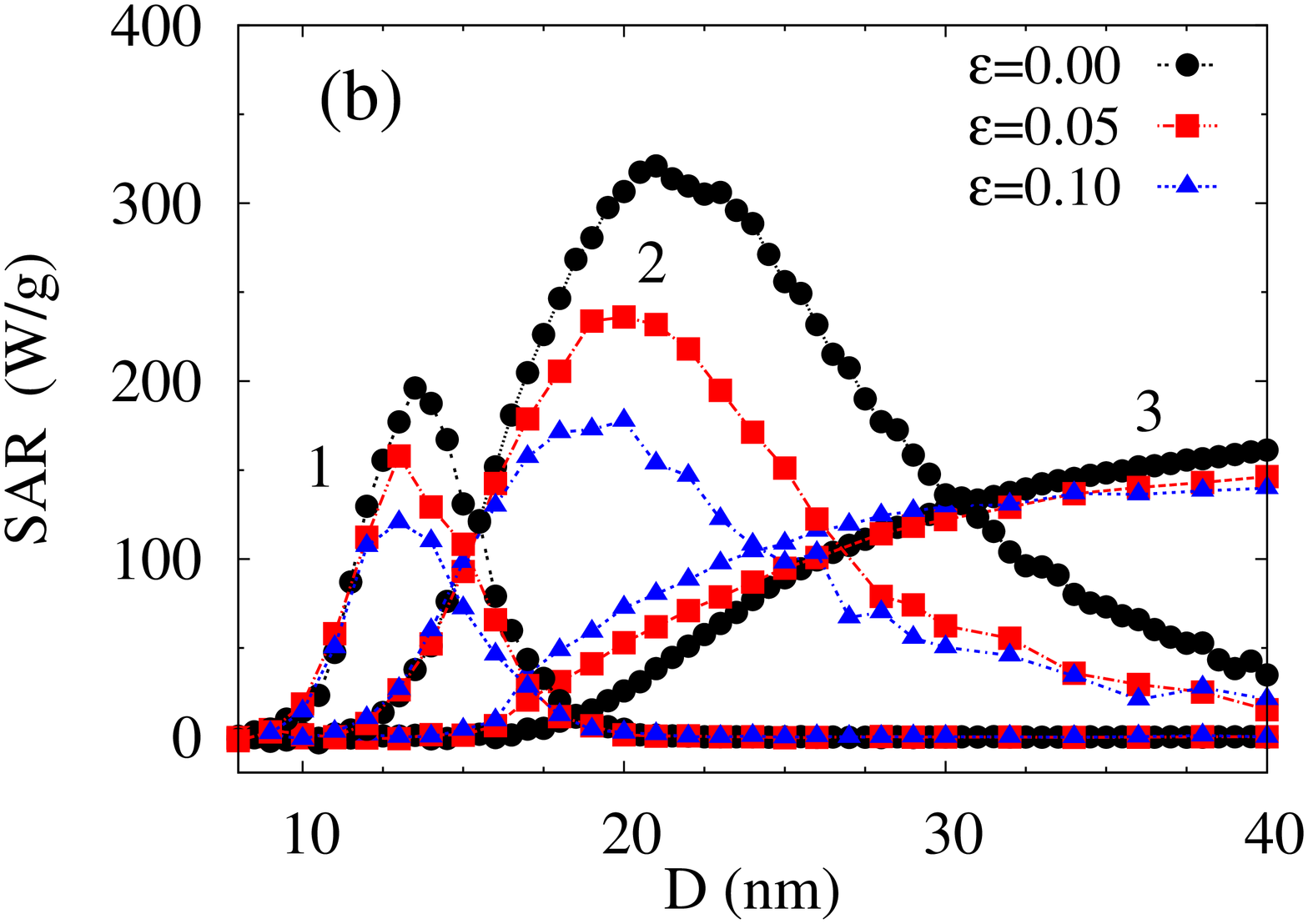}
\end{minipage}
\caption{SAR as a function of the mean particle size $D$ for variable volumetric packing fraction $\epsilon$ including non-interacting case ($\epsilon=0$) and two values $M_s=200 $ emu/cm$^3$  and 400 emu/cm$^3$. For low $M_s=200 $ emu/cm$^3$ the dipolar interactions are weak and SAR does not depend on $\epsilon$. The solid line represents the solutions by means of RT. (b): SAR vs. $D$ for different packing fractions and the values of $K = 3 \times 10^5$ erg/cm$^3$ (curve set 1),  $1.5 \times 10^5$ erg/cm$^3$ (curve set 2) and $0.5 \times 10^5$ erg/cm$^3$ (curves set 3). Calculations correspond to log-normal distributions of size and anisotropy constants both with standard deviations 0.1, spherically distributed anisotropy axes, and field amplitude $H_0 = 300$ Oe.}
\label{fig11}
\end{figure*}

\subsection{Importance of dipolar interactions}
Next we investigate the effect of dipolar interactions, which have been ignored in the study so far. For this we consider a system created by the random positioning of particles in a 3D configuration, as illustrated in Fig. \ref{fig12}(c). Dipolar interactions are truncated and handled within the minimum image convention imposed by the periodic boundary conditions. We verified that the truncation does not affect the radial correlation function in the relevant model parameter range and thus does not modify any essential physics. We choose particle assemblies with volumetric packing fractions from the interval $\epsilon = 0.0, \dots, 0.14$, which gives another way of controlling dipolar effects on SAR in the particle assembly in this work besides modifying the value of $M_s$.

The dependence of SAR on the mean particle diameter for two different values of $M_s$ is shown in Fig. \ref{fig11}(a). For low $M_s = 200$ emu/cm$^3$ the dipolar interactions are weak in comparison to the chosen anisotropy $K$ (i.e. low $\zeta ( = M_{s}^{2}\epsilon/K) $), and as a result changing the packing fraction $\epsilon$ does not have any significant effect on SAR. The reduced value of $M_s$ decreases the initial susceptibility $\chi_0$ in Eq. \eqref{rosensweig} and thus the overall SAR. It also increases $H_K$, which lowers the ratio $H_0/H_K$ leading to the suppression of the overall nonlinear behaviour and improvement of the accuracy of RT.
On the other hand, for $M_s = 400$ emu/cm$^3$ interactions begin to play a significant role and SAR becomes dependent on $\epsilon$ (Fig. \ref{fig11}(a)). Given the assumed spherical distribution of anisotropy axes, the effect of dipolar interactions is to reduce the energy barriers separating the minimum energy states \cite{Hovorka2014}, and thus to lower the effective particle anisotropy in comparison to the non-interacting case value $K$. The $\chi_0$ and the N\'eel relaxation time constant $\tau_N$ (relative to field variation time scale $f^{-1}$) become smaller with increasing $\epsilon$, which according to Eq. \eqref{rosensweig} reduces SAR and shifts its peak towards smaller $D$. 

In Fig. \ref{fig11}(b) these calculations are further extended toward the hysteretic regime by decreasing the value of $K$, similarly to Fig. 3,
indicating that, in this regime the effect of interactions is more complex. In particular the trend can change qualitatively for small values of $K$, where increasing $\epsilon$ can even lead to enhancement of SAR (the set of curves 3). The reason is, that the value of $K$  essentially determines the dominant mechanism of heating. When $K$ is large, the heating mechanism is predominantly of superparamagnetic nature and generated via dynamic hysteresis, and here the interactions are seen to decrease the SAR value. 
At the opposite limit of low values of $K$, the external field amplitude is sufficient to cause hysteretic switching of particles in the system, and as a result the heating mechanism is predominantly via conventional metastable hysteresis. In this case, increasing $\epsilon$ results in the increasing SAR when particle diameter is small while for larger particle the trend is opposite.  This behaviour is consistent with previous model calculations of Verdes  \emph{et. al.}\cite{Verdes2002}, who showed that the effects of interactions on the hysteretic properties are indeed dependent on the value of $KV$ via the parameter $\alpha = KV/kT$. For small $\alpha$ i.e., close to superparamagnetic systems, the interactions tend to be pairwise, leading to an enhancement of the energy barrier, whereas for hysteretic systems (large $\alpha$) the microstructure is dominated by a tendency to flux closure leading to energy barrier reduction.

\begin{figure}[h!]
\begin{center}
\hspace{0.04\textwidth}
\includegraphics[angle = -90,width=\linewidth]{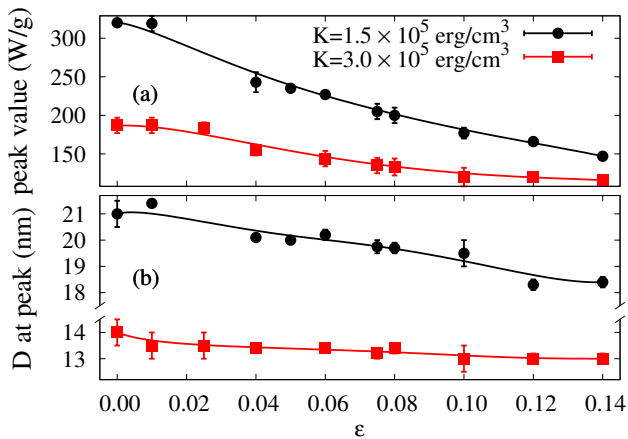}
\caption{ Maximum SAR value (a) and peak position (b) as function packing fraction, $\epsilon$ for two values of anisotropy: 3 $ \times 10^5 erg/cm^3$ (red squares) and  1.5 $ \times 10^5 erg/cm^3$ (black circles). The lines are to guide the eye.  }
\label{fig_peakandpos}
\end{center}
\end{figure}
Fig. \ref{fig_peakandpos} demonstrates the important effect of interactions, via the dependence of the maximum SAR value and the optimal particle size as a function of the packing fraction $\epsilon$. Consistently with Fig. 4, the interaction strength scales with $\zeta$, and the peak value of SAR clearly decreases with increasing interaction. The decrease of the interaction effects for the  larger $K$ values is also consistent with this scaling. Clearly, the effect of interactions on the SAR, and also on the optimal particle diameter, should not be neglected. We note that these calculations are carried out for relatively low particle densities: for systems containing aggregated particles, where the local density can be significantly higher, interaction effects may be even more pronounced \cite{Serantes2010,Branquinho2013,Serantes2014}.

\section{Discussion}
\begin{figure*}[ht!]
\begin{center}
\hspace{0.04\textwidth}
\includegraphics[angle = -90,width=\linewidth]{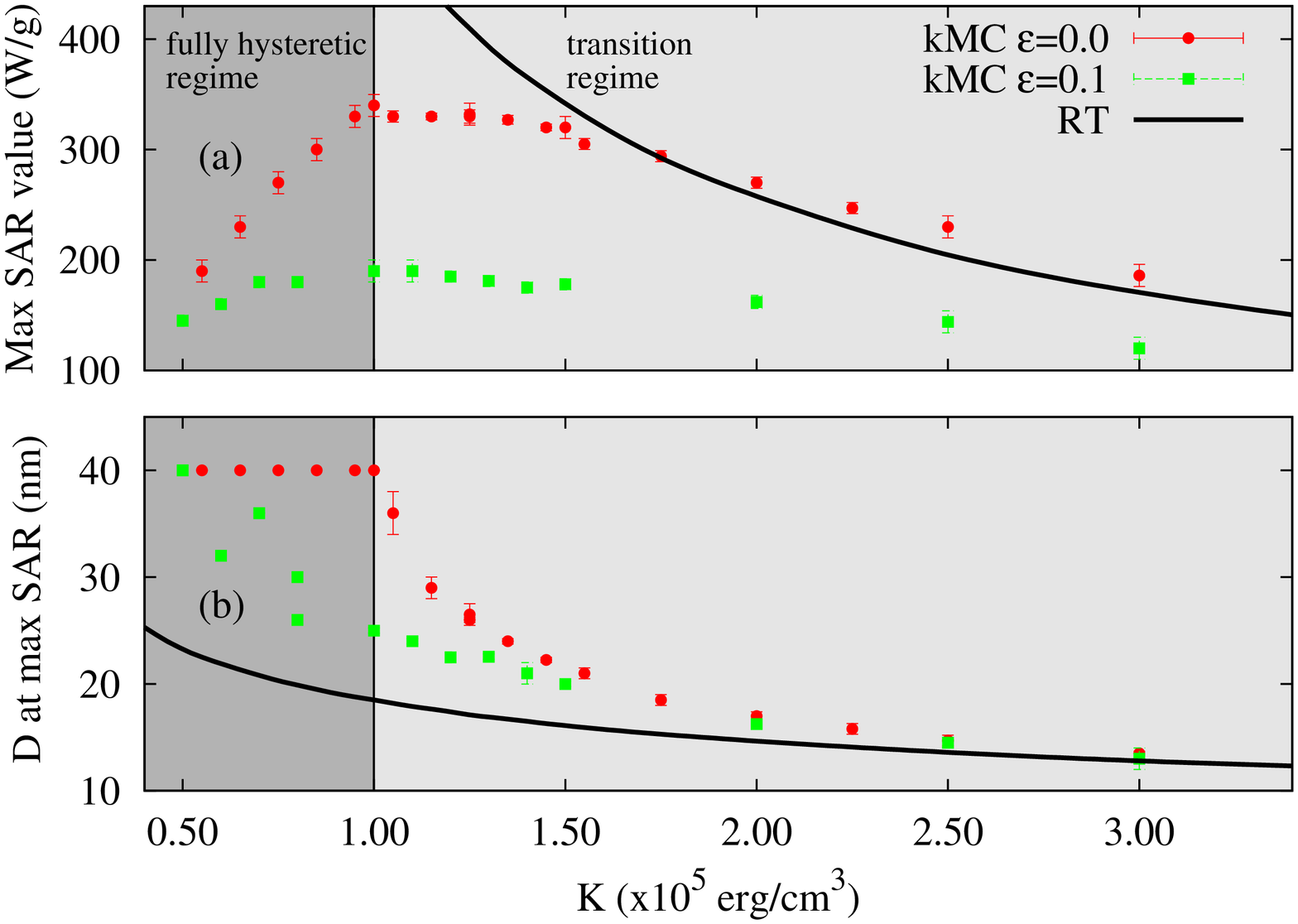}
\caption{ Maximum SAR value (a) and the corresponding particle size (b) as function of anisotropy for: RT (black line), kMC non-interacting case (red circles) and kMC with interaction (green squares). In the fully hysteretic regime the SAR dependence on D does not present a peak, but is reaching a saturation value with increasing D.  }
\label{fig_K_opt_SAR}
\end{center}
\end{figure*}

We have developed a kinetic Monte-Carlo model of the underlying heating mechanisms associated with the hyperthermia phenomenon used in cancer therapy. 
Our work shows that the apparently different (superparamagnetic and hysteretic) mechanisms of magnetic heating have a common, hysteretic, physical origin. The so-called superparamagnetic regime is characterised by a dynamic hysteresis which transforms gradually to more conventional hysteresis arising from thermally activated transitions between metastable states.
The kMC model described here captures the physics of  both heating mechanisms. In the linear response regime, the kMC results are in agreement with the theory of Rosensweig~\cite{Rosensweig2002}. Beyond this limit the nonlinear contributions, included in the kMC approach, give rise to increasing divergence from the predictions of Ref. \onlinecite{Rosensweig2002}, leading to significantly different SAR values and to different materials parameters for optimal heating (Fig. \ref{fig_K_opt_SAR}). The transition regime is characterise by two main features: 1) large values of SAR and 2) SAR dependence on D presents broad peaks. Both characteristics are ideal for hyperthermia application, therefore making the transition region of high importance.

We also carried out an investigation of the effect of dipolar interactions. The effects of interactions were found to be significant and also generally consistent with a scaling of the relative interaction strength with $\zeta$ (Fig. \ref{fig_peakandpos}). However,  the effects of interactions are also dependent on the ratio $KV/k_BT$ since the interactions affect the properties of superparamagnetic and hysteretic systems differently as reported in Ref. \onlinecite{Verdes2002}.  Interactions are rarely considered in relation to models of hyperthermia, but the results presented here suggest that their inclusion, along with detailed models of the nanoparticle structures, are extremely important for realistic predictions of the heat generation which is central to the phenomenon.

Both the non-linear behaviour induced at higher applied field amplitudes $H_0$ and the dipolar interactions produce a significant deviation from the RT. Figure \ref{fig_K_opt_SAR} illustrates the complexity of the hyperthermia problem. The variation of the optimum SAR value as function of K is not monotonic. In the transition regime the optimum SAR value increases with decreasing K. Afterwards, when entering in the fully hysteretic regime, SAR decreases with decreasing K. In this regime, SAR vs D does not present a peak, but asymptotically reaches a saturation value with increasing size. In this case after a certain size there is no significant variation in SAR. This is shown in Fig. \ref{fig_K_opt_SAR}b. Considering now the interacting case, the transition towards the fully hysteric regime is not as clear as the non-interacting case. Although the maximum SAR decreases with decreasing K, the particle size for reaching the maximum SAR is varying.

In conclusion, the kMC approach gives a unified treatment of the apparently disparate mechanisms of heating in the superparamagnetic and fully hysteretic regimes and the transition between them and, with its accurate timescale quantification, gives a reliable prediction of the frequency and particle size dependent behaviour. The inclusion of interparticle interactions leads to important and complex effects on the hysteresis which will be amplified in systems involving aggregated particles. Further development and long-term optimisation of the hyperthermia phenomenon must be cognisant of these factors and the modelling approach described here is a powerful tool for future materials and therapy design.

\section{Methods}

\subsection {kMC model}
The basic model, which follows Ref. \onlinecite{Chantrell2000}, is a collection of $N$ Stoner-Wohlfarth spherical particles $i$ with uniaxial anisotropy $\vec k_i$ and diameter $D_i$ confined in sample volume $V_{s}$. The interaction field coupling any pair of particles $i$ and $j$ reads:
\begin{equation}
\vec H_i^\textrm{dip} = \sum_{i\ne j}V_jM_sr_{ij}^{-3}(-\hat m_j +  3\hat r_{ij}(\hat m_j\cdot\hat r_{ij}))
\label{dipolefield}
\end{equation}
Where $\vec r_{ij} = r_{ij}\hat r_{ij}$ is the inter-particle distance. The volumetric packing fraction of particles is $\epsilon=N \langle V_i \rangle /V_s = NV/V_s$ and the average interparticle distance is $l =(V_{s}/N)^{1/3} $. The system is characterised by the dipole energy ($\sim M_{s}^{2}  V  ^2/l^3$), anisotropy energy ($K  V$) and thermal energy ($kT$). The effect of interactions will depend on the  thermal stability ($\alpha=KV/kT$) and the ration between dipole energy and anisotropy energy ($ \zeta =M_{s}^{2} V^2 /(l^3 K V)= M_{s}^{2}\epsilon/K$).

Superparamagnetic  behaviour occurs up to large energy barriers. Persistence in the superparamagnetic behaviour creates difficulties for standard MC approaches due to the unreasonably large number of MC steps that are necessary to achieve equilibrium. By considering the superparamagnetic particles with large energy barriers($>3kT$) as a two-state system, an improved computational approach can therefore be derived \cite{Chantrell2000}.
For effective local fields of particles $\vec H+\vec H_i^\textrm{dip}$ less than a critical value there are two equilibrium particle states, `$+$' and `$-$', and the probability for a particle moment $\hat m_i$ to switch between these states is $P_i = 1-\exp(-t/\tau_i)$ \cite{Hovorka2014}. The relaxation time constant $\tau_i$ is a reciprocal sum of the transition rates $\tau^+_i$ and $\tau^-_i$ dependent on the energy barriers $\Delta E_i^\pm$ seen from the `$+$' and `$-$' states via the standard N\'eel-Arrhenius law\cite{Neel1949}: $\tau_i^\pm = \tau_0\exp(\Delta E_i^\pm/k_BT)$, where $k_B$ is the Boltzmann constant and $T$ the temperature. The method has a real time step due to the master-equation nature and can be applied for frequency up to $10^7 Hz$. There is no low  frequency limit for the method. 

\subsection {Generating particle spatial configuration}
To obtain a system of random particles configuration having a certain packing fraction, $\epsilon$, we start with a perfect simple cubic lattice with a  large lattice spacing so that the particles do not overlap. Then the particles are randomly moved  with a classical Metropolis Monte-Carlo approach. The move is accepted with the probability $\min (1,e^{(-\Delta E)})$. This step is repeated 50 times and then the system size and the inter-particle distance are reduced with an amount, for which there are no touching particles.   The procedure is repeated  until the desired packing fraction is obtained. Afterwards 500 more random moves are done for each particle. $\Delta E$ is calculated using a dimensionless repulsive energy of the form $E_{i}^{new,old}=\sum_{\substack{j=1} , j\ne i}^{N} {1000 \cdot D_i/({r_{ij}^{new,old}})^4 } $.

\subsection {Rosensweig theory}
To compare and validate our kMC model with the Rosensweig theory (RT) we used all possible consideration for improving the original theory.
For the equilibrium susceptibility $\chi_0$  we used the equilibrium magnetization curves calculated as in Ref. \onlinecite{Chantrell2000}, which incorporates the dependence on the distributions of particle size, anisotropy values, and anisotropy easy axis directions. In this case Eq. \ref{rosensweig} becomes:

\begin{equation}\label{rosensweig_updated}
 {\cal P} = \frac{1}{N} \sum_{i=1}^{N} {\pi\mu_0\chi_{0}(\vec k_i,D_i) H_0^2 f\frac{2\pi f\tau_i(K_i,D_i)}{1+(2\pi f\tau_i(K_i,D_i))^2}}
\end{equation}

Eq. \ref{rosensweig_updated}  allows to calculate the specific heat power for any system of non-interacting magnetic particles. For example for the ideal case when the easy axis is parallel with the field amplitude. 

We calculate hysteresis curves numerically using the kMC model and from the expression arising from RT: $M(H(t))=H_0 ( \chi^{'} \cos(2 \pi f t) + \chi^{''} \sin(2 \pi f t) )$. This represents the equation of an ellipse, with $\chi^{'}$ and $\chi^{''}$ calculated from Eq. \eqref{master} using the linear response theory:
\begin{align}
\chi^{'}&=\frac{\chi_0}{1+(2\pi f \tau)^2}\\
\chi^{''}&=\frac{\chi_0}{1+(2\pi f \tau)^2} (2 \pi f \tau)
\label{eq1}
\end{align} 

\subsection{Evaluating SAR}
 Denoting the hysteresis loop area as $A$, the specific heating power is computed in simulations directly as the product ${\cal P} = Af M_s V_t$, which is then compared with calculations based on Eq. \eqref{rosensweig} as described below. The quantity that characterises the efficiency of the heating is known  as the specific absorption rate (SAR) or heat dissipation per unit mass: SAR = ${\cal P}/m_t= Af M_s/\rho$; where $ \rho$ is the density of magnetic material, $V_t$/$m_t$ is the total volume/mass of magnetic material. In this paper we consider $\rho=5.2 g/cm^3$, the density of magnetite (\it{${Fe_3 O_4}$}).


\begin{thebibliography}{1}
\bibitem{dynvsmeta}
It is worthwhile pointing out that the $M(H)$ behavior in the dynamical
  response theory regime resembles a hysteresis loop. However, this kind of
  hysteresis is of dynamical character and essentially associated with
  `inertial' phase lag between $M$ and $H$, and is fundamentally different from
  the type of hysteresis associated with metastability.
\end{thebibliography}
\end{document}